# An $L_0L_1$-norm compressive sensing paradigm for the construction of sparse predictive lattice models using mixed integer quadratic programming

Wenxuan Huang[1,3], Alexander Urban[2], Penghao Xiao[3], Ziqin Rong[1,3], Hena Das[2], Tina Chen[2], Nongnuch Artrith[2], Alexandra Toumar[1,3], Gerbrand Ceder[1,2,3]

1, Department of Material Science and Engineering, Massachusetts Institute of Technology, MA, USA

2, Department of Materials Science and Engineering, UC Berkeley, Berkeley, CA, USA

3, Materials Science Division, Lawrence Berkeley National Laboratory, Berkeley, CA, USA

## Abstract

First-principles based lattice models allow the modeling of *ab initio* thermodynamics of crystalline mixtures for applications such as the construction of phase diagrams and the identification of ground state atomic orderings. The recent development of compressive sensing approaches for the construction of lattice models has further enabled the systematic construction of sparse physical models without the need for human intuition other than requiring the compactness of effective cluster interactions. However, conventional compressive sensing based on $L_1$-norm regularization is strictly only applicable to certain classes of optimization problems and is otherwise not guaranteed to generate optimally sparse and transferable results, so that the method can only be applied to some materials science applications. In this paper, we illustrate a more robust $L_0L_1$-norm compressive-sensing method that removes the limitations of conventional compressive sensing and generally results in sparser lattice models that are at least as predictive as those obtained from $L_1$-norm compressive sensing. Apart from the theory, a practical implementation based on state-of-the-art mixed-integer quadratic programming (MIQP) is proposed. The robustness of our methodology is illustrated for four different transition-metal oxides with relevance as battery cathode materials: $Li_{2x}Ti_{2(1-x)}O_2$, $Li_{2x}Ni_{2y}O_2$, $Mg_xCr_2O_4$, and $Na_xCrO_2$. This method provides a practical and robust approach for the construction of sparser and more predictive lattice models, improving on the compressive sensing paradigm and making it applicable to a much broader range of applications.

## Introduction

*First principles* electronic density-functional theory (DFT) [1,2] has enabled predictive computational material science [3-9] and serves as an important tool for discovering novel materials [3,4,10]. However, the intrinsic computational scaling of DFT limits its applicability in practice to structures with a few hundred atoms and typically prevents the extensive configurational sampling that is required to examine temperature-dependent properties. To overcome this limitation, *lattice models* based on previously calculated DFT energies, such as *cluster expansion* (CE) or generalized Ising models [11-16], can be used to efficiently approximate first-principles energies of lattice configurations that are out of reach for DFT calculations.

As the name suggests, CE models approximate the energies of lattice configuration as a linear expansion in the occupancy (*effective cluster interactions*, ECIs) of site clusters, *i.e.*, site pairs, triplets, quadruplets, and so on. Typically, a small number of short-ranged clusters with few sites is sufficient to capture the characteristic physics of a material. However, it is not normally known *a priori* which clusters correspond to the most important interactions. Constructing a cluster expansion model, therefore, involves the solution of an underdetermined linear system in the cluster basis, with the objective of creating an optimal solution which is sparse, so that most ECIs are zero and only the ECIs of the characteristic clusters contribute to the CE energy.

Traditionally, the characteristic clusters for a material were determined either based on human intuition [17] or using global optimization techniques, such as genetic algorithms [18]. Recently, important progress in CE model construction has been made [19-23] by making use of compressive sensing [24], a signal reconstruction technique that finds the sparsest solution for a certain type of underdetermined linear systems. The compressive-sensing CE approach allows the systematic and robust determination of sparse ECI parameters without the need for human intuition other than requiring the compactness of ECIs. However, the conventional $L_1$-norm compressive sensing is only guaranteed to determine the optimally sparse solution of linear systems that fulfill a number of conditions, in particular the restricted isometry property (RIP) [25]. In practical material science and physics applications, these properties do not necessarily hold, and applying $L_1$-norm compressive sensing does not automatically lead to the optimal model.

In this paper, we explore a generalization of the conventional $L_1$-norm compressive sensing tailored for the domain of practical materials science CE problems that is not bound to the stringent requirements of the original method. We demonstrate that our $L_0L_1$-norm compressive sensing method leads to significantly sparser and more predictive models.

The paper is structured as follows: First, we revisit the CE formalism and introduce the ECI fitting problem mathematically in the *methodology* section. Then, we introduce the $L_0L_1$-norm compressive sensing paradigm and its mixed integer programming formulations, which enable efficient and near optimal solution of the relevant *nondeterministic polynomial time* (NP) hard problems with state of the art performance. Finally, in the *results* section we demonstrate that the $L_0L_1$-norm formulation enables significantly sparser and even more predictive CE models compared to the conventional $L_1$-norm formulation.

As a side note, ensuring the preservation of ground states is another important challenge in lattice model construction which has been addressed in reference [26]. The present paper focuses exclusively on constructing sparser lattice models but the formalism is compatible with that of reference [26] and should be understood as complementary

## Methodology

### Cluster expansion and compressive sensing

In this section, we briefly recap the concepts of cluster expansion. For more elaborate introductions, the readers are referred to the literature [11,14,19,27].

The general expression of a cluster expansion Hamiltonian is

$$\mathbf{E}_{CE}(\sigma) = \left\langle J_0 + \sum_i^{sites} J_i \sigma_i + \sum_{i,j}^{pairs} J_{i,j} \sigma_i \sigma_j + \sum_{i,j,k}^{triplets} J_{i,j,k} \sigma_i \sigma_j \sigma_k + ... \right\rangle = \sum_{c \in \mathbf{C}} J_c \left\langle \sigma_c \right\rangle = \Pi(\sigma)\mathbf{J}, \quad (1)$$

where $\sigma$ is the lattice (*spin*) configuration with $\sigma_i$ denoting the occupancy of site *i*. Conventionally, $\sigma_i$ is +1 when the site is *spin up* in a magnetic system and -1 when the site is *spin down* though the variables can denote any binary occupancy. The polynomial $\sigma_i \sigma_j ...$, denotes the *spin product* of a *cluster* of lattice sites, associated with the weight factor $J_{i,j...}$, the *effective cluster interaction* (ECI). The CE energy is then the sum of the spin products multiplied with their corresponding ECIs and averaged over all sites within a lattice, as denoted by the angle brackets $\langle \rangle$. To simplify notation, with $\mathbf{C}$ defined as the set of clusters, for any $c \in \mathbf{C}$, $J_c$ and $\sigma_c$ are shorthand notations of the previously defined ECIs $J_{i,j...}$ and spin products $\sigma_i \sigma_j ...$, respectively. The average of the spin products $\langle \sigma_c \rangle$ is the *cluster correlation* of cluster *c*. By concatenating the cluster correlations $\langle \sigma_c \rangle$ into a row vector $\Pi(\sigma)$ and the ECIs $J_c$ into a column vector $\mathbf{J}$, the CE energy of the lattice configuration $\sigma$ can be simply written as the vector product $\Pi(\sigma)\mathbf{J}$.

Naïvely, the CE fitting problem can be formulated as follows: Given a set of input configurations $S$ and their energies from DFT calculations $\mathbf{E}_{\text{DFT},S}$, minimize the root mean square error (RMSE) between the CE energy and DFT energy in the ECIs space, *i.e.*, $\min_{\mathbf{J}} \|\mathbf{E}_{\text{DFT},S} - \mathbf{E}_{\text{CE},S}\|_2$. By concatenating the correlation vectors of all configurations $\sigma \in S$ we obtain the correlation matrix $\Pi_S$, which allows us to write more explicitly $\mathbf{E}_{\text{CE},S} = \Pi_S \mathbf{J}$, so that the naïve statement of the CE fit problem becomes:

$$\min_{\mathbf{J}} \|\mathbf{E}_{\text{DFT},S} - \Pi_S \mathbf{J}\|_2^2 . \qquad (2)$$

For $p \geq 1$, the $L_p$ norm of a vector $\mathbf{u}$ is defined as follows:

$$\|\mathbf{u}\|_p = \left(\sum_i |u_i|^p\right)^{1/p} \quad \text{if} \quad p \geq 1 . \qquad (3)$$

The $L_0$ norm ($p = 0$) will be used extensively in this paper:

$$\|\mathbf{u}\|_0 = \sum_i \text{Ind}(u_i) , \qquad (4)$$

where $\text{Ind}(\bullet)$ is a binary indicator function:

$$\text{Ind}(u_i) = \begin{cases} 1 & \text{if } u_i \neq 0 \\ 0 & \text{if } u_i = 0 \end{cases} . \qquad (5)$$

Generally speaking, the $L_0$ norm measures the sparseness of a vector by counting the number of non-zero elements in the vector. However, it is worthwhile to mention that the "$L_0$ norm" is, despite the terminology, not formally a norm since the basic property of *homogeneity of norm* is generally not satisfied, i.e., $\|a\mathbf{u}\|_0 \neq |a|\|\mathbf{u}\|_0$.

In practice, the dimension of $\mathbf{J}$ is generally larger than the number of available DFT reference calculations, *i.e.*, the equation system is underdetermined, so that Eq. (2) possesses an infinite number of solution vectors with an RMSE of 0. The direct solution of Eq. (2) therefore results in *over-fitting*, where the in-sample data is perfectly reproduced at the cost of losing predictive power on out-of-sample data. A typical result of over-fitting in CE is that the ECI vector, $\mathbf{J}$, has many non-zero elements of large magnitude but with opposite signs. To avoid over-fitting, compressive sensing [19,28] adds $L_1$ norm regularization to the fitting procedure, resulting in the objective function

$$\min_{\mathbf{J}} \left\| \mathbf{E}_{DFT,S} - \mathbf{\Pi}_S \mathbf{J} \right\|_2^2 + \mu_1 \left\| \mathbf{J} \right\|_1 \tag{6}$$

where $\mu_1 > 0$ is the $L_1$ norm regularization constant. Candès *et al.* proved that minimizing the $L_1$ norm is, for many problems, equivalent to minimizing the $L_0$ norm [25], so that the $L_1$ norm regularization term $\mu_1 \left\| \mathbf{J} \right\|_1$ is essentially a penalty that forces $\mathbf{J}$ to be sparse with components of relatively small magnitude, reducing over-fitting and allowing the construction of a robust CE model. However, the proof of reference [25] requires that the correlation matrix $\mathbf{\Pi}_S$ satisfies the restricted isometry property (RIP) [25] for Eq. (6) to provide the most accurate and sparse solution. In essence, RIP requires $\mathbf{\Pi}_S$ to be close to orthonormal (every principle submatrix of $\mathbf{\Pi}_S^T \mathbf{\Pi}_S$ to be close to identity matrix).

As discussed in reference [19], to satisfy RIP in the context of CE models, the reference lattice configurations in $\mathbf{\Pi}_S$ could be drawn such that the correlations cover the correlation space randomly. However, many relevant physical properties, such as thermodynamic phase diagrams, depend on highly accurate low-energy states but tolerate larger errors for lattice configurations with high energies. From that perspective, $\mathbf{\Pi}_S$ should preferably contain many more low-energy states than high energy states as it is more important that the CE model is able to distinguish between low energy states. Unfortunately, such a non-randomly sampled $\mathbf{\Pi}_S$ does not generally satisfy the RIP, and solving Eq. (6) directly may not lead to optimal results.

### *L₀L₁* normalization

The goal of this paper is, thus, to develop a CE construction based on a compressive sensing paradigm without the RIP assumption. The original compressive sensing CE paradigm aims at using $L_1$ norm regularization to reproduce the results of $L_0$ norm regularization, as $L_1$ norm regularization can be solved computationally efficiently using quadratic programming techniques [26]. Our proposed approach is to reintroduce $L_0$ norm regularization as the exact measure of sparseness into Eq. (6) to obtain a sparser and more accurate solution. In other words, the objective function becomes

$$\min_{\mathbf{J}} \left\| \mathbf{E}_{DFT,S} - \mathbf{\Pi}_S \mathbf{J} \right\|_2^2 + \mu_1 \left\| \mathbf{J} \right\|_1 + \mu_0 \left\| \mathbf{J} \right\|_0 \tag{7}$$

with $\mu_0 > 0$ being the $L_0$ norm regularization constant. The $L_1$ norm term in Eq. (7) is retained to assists the $L_0$ norm regularization, as the solution of $L_1$ norm minimization is already a close approximate to the exact solution of Eq. (7). Thus, intuitively speaking, the $L_1$ norm term guides the search for a better $L_0$ norm regularization result. The downside of this construction is that the $L_0$ norm minimization problem in Eq. (7) is NP-hard [29]. Practically, this means no algorithm can guarantee to obtain its exact solution in polynomial time. However, our main goal is to improve the CE model construction results beyond the conventional $L_1$ norm regularization, and any improved solution to Eq. (7) is desirable as long as it can be obtained within practical time limits.

### Mixed integer quadratic programming for *L₀L₁*-norm compressive sensing

To implement CE model construction based on the objective function of Eq. (7), we make use of a mathematical programming [30] technique called *mixed integer quadratic programming* (MIQP) [31,32]. In recent years, important developments have been made in heuristic methods [33] for the solution of MIQP without proof of optimality, including 1-opt, 2-opt [34], local branching [35], Relaxation Induced

Neighborhood Search (RINS) [36] and polishing [37] based on evolutionary algorithms. These heuristics enable the practical and efficient search for near-optimal solutions even if the underlying problem is NP-hard. Exact methods (with proof of optimality) including branch-and-bound [38,39], cutting planes [40,41] and branch-and-cut [42,43] have also been developed extensively. Building on these theoretical developments, there have been immense efforts to develop robust and efficient state-of-the-art solvers such as Gurobi [44] that fine-tune and integrate different heuristics and exact methods.

To solve the $L_0L_1$-norm compressive sensing problem of Eq. (7), we convert Eq. (7) into an MIQP problem and use Gurobi to solve for its near optimal solution within a practical time span. The precise definition of MIQP and the conversion of Eq. (7) into an MIQP problem are elaborated in the following.

MIQP is a mathematical optimization technique for problems of the standard form:

$$\begin{aligned} \min_{\mathbf{x}} \quad & \frac{1}{2}\mathbf{x}^T\mathbf{Q}\mathbf{x} + \mathbf{c}^T\mathbf{x} \\ s.t. \quad & \mathbf{A}\mathbf{x} \leq \mathbf{b} \\ & \mathbf{C}\mathbf{x} = \mathbf{d} \\ & x_i \in \mathbb{Z} \qquad \forall i \in I \end{aligned} \qquad (8)$$

where $\mathbf{Q}$ is a positive semi-definite matrix, i.e., for all real vectors $\mathbf{x}$, $\mathbf{x}^T\mathbf{Q}\mathbf{x} \geq 0$. The positive semi-definite property ensures that the function to be optimized is convex. $\mathbf{A}$ and $\mathbf{C}$ are real matrices, and $\mathbf{b}$, $\mathbf{c}$, and $\mathbf{d}$ are real vectors. $\mathbb{Z}$ is the set of integers and $I$ is a set of indices for $\mathbf{x}$ to be integers. Note that replacing $\mathbb{Z}$ with $\{0,1\}$ leads to an important subclass of MIQP [45], due to the relationship:

$$x_i \in \{0,1\} \quad \Leftrightarrow \quad x_i \in \mathbb{Z} \text{ and } 0 \leq x_i \leq 1. \qquad (9)$$

We convert Eq. (7) to an MIQP by expanding the quadratic term:

$$\min_{\mathbf{J}} \mathbf{J}^T \Pi_S{}^T \Pi_S \mathbf{J} - 2\mathbf{E}_{\text{DFT},S}{}^T \Pi_S \mathbf{J} + \mathbf{E}_{\text{DFT},S}{}^T \mathbf{E}_{\text{DFT},S} + \mu_1 \|\mathbf{J}\|_1 + \mu_0 \|\mathbf{J}\|_0 . \qquad (10)$$

Note that the term $\mathbf{E}_{DFT,S}{}^T\mathbf{E}_{DFT,S}$ just gives rise to a constant shift that has no effect on the optimal solution $\mathbf{J}$. $\Pi_S{}^T\Pi_S$ is positive semi-definite because $\mathbf{x}^T\Pi_S{}^T\Pi_S\mathbf{x} = \|\Pi_S\mathbf{x}\|_2^2 \geq 0$. The key step to further convert Eq. (10) into a MIQP is to introduce auxiliary variables $z_{0,c}$ and $z_{1,c}$ for each cluster $c \in \mathbf{C}$ to track $\mathrm{Ind}(J_c)$ and $|J_c|$ respectively. Explicitly, Eq. (10) is equivalent to:

$$\min_{\mathbf{J},\mathbf{z}} \mathbf{J}^T\Pi_S{}^T\Pi_S\mathbf{J} - 2\mathbf{E}_{DFT,S}{}^T\Pi_S\mathbf{J} + \mathbf{E}_{DFT,S}{}^T\mathbf{E}_{DFT,S} + \mu_1\sum_{c\in\mathbf{C}}z_{1,c} + \mu_0\sum_{c\in\mathbf{C}}z_{0,c}$$

$$s.t. \quad z_{1,c} \geq J_c \quad \forall c \in \mathbf{C}$$
$$z_{1,c} \geq -J_c \quad \forall c \in \mathbf{C}$$
$$Mz_{0,c} \geq J_c \quad \forall c \in \mathbf{C} \quad , \quad (11)$$
$$Mz_{0,c} \geq -J_c \quad \forall c \in \mathbf{C}$$
$$z_{0,c} \in \{0,1\} \quad \forall c \in \mathbf{C}$$

where $M$ is some large number which $|J_c|$ should never exceed for every cluster $c \in \mathbf{C}$. For example, $M$ could be set to 50 *eV/f.u.* (electronvolt per formula unit) as long as we know intuitively that the magnitude of ECIs should never exceed 50 *eV/f.u.* Inclusion of such $M$ parameters is generally referred to as "big $M$ formulation" in mathematical programming [30].

A proof of the equivalence between Eq. (10) and Eq. (11) is provided in the following section.

**Proof of the equivalence of the $L_0L_1$-norm regularized cluster expansion objective function Eq. (10) and its MIQP reformulation Eq. (11)**

To prove the equivalence of Eq. (10) and Eq. (11), we show:

**(A)** For any optimal solution $(\mathbf{J}',\mathbf{z}_1',\mathbf{z}_0')$ to Eq. (11), $\mathbf{J} = \mathbf{J}'$ is an optimal solution to Eq. (10).

**(B)** For any optimal solution $\mathbf{J}$ to Eq. (10), $\mathbf{J}' = \mathbf{J}$, $z'_{1,c} = |J_c|$ and $z'_{0,c} = \text{Ind}(J_c)$, for all $c \in \mathbf{C}$ is an optimal solution to Eq. (11).

An important lemma (L1) needs to be established before the proof of (A), (B).

**(L1)** $(\mathbf{J}', \mathbf{z}'_1, \mathbf{z}'_0)$ is optimal to Eq. (11) implies $z'_{1,c} = |J'_c|$ and $z'_{0,c} = \text{Ind}(J'_c)$ $\forall c \in \mathbf{C}$.

**Proof of (L1):**

(i) $z'_{1,c} = |J'_c|$ $\forall c \in \mathbf{C}$. If $z'_{1,c} < |J'_c|$, it violates either $z'_{1,c} \geq J'_c$ or $z'_{1,c} \geq -J'_c$. If $z'_{1,c} > |J'_c|$, the objective function in Eq. (11) is strictly greater than for $z'_{1,c} = |J'_c|$ since $\mu_1 > 0$. Therefore $z'_{1,c} = |J'_c|$.

(ii) $z'_{0,c} = \text{Ind}(J'_c)$ $\forall c \in \mathbf{C}$. Note that $z'_{0,c} \in \{0,1\}$ and $\text{Ind}(J'_c) \in \{0,1\}$. Suppose $z'_{0,c} = 0$ and $\text{Ind}(J'_c) = 1$. We know that $J'_c \neq 0$ (since $\text{Ind}(J'_c) = 1$). However, the required constraints $Mz'_{0,c} = 0 \geq J'_c$ and $-Mz'_{0,c} = 0 \leq J'_c$ would lead to $J'_c = 0$. This is a contradiction and thus $\text{Ind}(J'_c) = 1$ implies $z'_{0,c} = 1$. Suppose $z'_{0,c} = 1$ and $\text{Ind}(J'_c) = 0$. We know that $J'_c = 0$, since $\text{Ind}(J'_c) = 0$. Taking $z'_{0,c} = 0$ leads to a strictly lower objective function value in Eq. (11) and no constraints are violated. This also leads to a contradiction since some parameters lead to a lower objective function value than $(\mathbf{J}', \mathbf{z}'_1, \mathbf{z}'_0)$. Therefore, $\text{Ind}(J'_c) = 0$ implies $z'_{0,c} = 0$. We thus conclude that $z'_{0,c} = \text{Ind}(J'_c)$. Therefore, if $(\mathbf{J}', \mathbf{z}'_1, \mathbf{z}'_0)$ is optimal for Eq. (11), this implies that $z'_{1,c} = |J'_c|$ and $z'_{0,c} = \text{Ind}(J'_c)$ $\forall c \in \mathbf{C}$.

**Proof of (A):**

Suppose $(\mathbf{J}', \mathbf{z}'_1, \mathbf{z}'_0)$ is optimal to Eq. (11) but $\mathbf{J} = \mathbf{J}'$ is not optimal to Eq. (10). Then there must be some other $\mathbf{J}''$ that optimizes Eq. (10). This means Eq. (10) evaluated at $\mathbf{J}''$ is strictly less than Eq. (10) evaluated at $\mathbf{J}'$. From (L1), we know that $z'_{1,c} = |J'_c|$ and $z'_{0,c} = \text{Ind}(J'_c)$ $\forall c \in \mathbf{C}$. Therefore, Eq. (11) evaluated at

$\left(\mathbf{J}',\left(|J'_c|\right)_{c\in\mathbf{C}},\left(\text{Ind}(J'_c)\right)_{c\in\mathbf{C}}\right)$ is equal to Eq. (10) evaluated at $\mathbf{J}'$. Note that $\left(\mathbf{J}'',\left(|J''_c|\right)_{c\in\mathbf{C}},\left(\text{Ind}(J''_c)\right)_{c\in\mathbf{C}}\right)$ is feasible (satisfying all constraints) to Eq. (11), and Eq. (11) evaluated at $\left(\mathbf{J}'',\left(|J''_c|\right)_{c\in\mathbf{C}},\left(\text{Ind}(J''_c)\right)_{c\in\mathbf{C}}\right)$ is equal to Eq. (10) evaluated at $\mathbf{J}''$. This implies that Eq. (11) evaluated at $\left(\mathbf{J}'',\left(|J''_c|\right)_{c\in\mathbf{C}},\left(\text{Ind}(J''_c)\right)_{c\in\mathbf{C}}\right)$ is strictly less than Eq. (11) evaluated at $\left(\mathbf{J}',\left(|J'_c|\right)_{c\in\mathbf{C}},\left(\text{Ind}(J'_c)\right)_{c\in\mathbf{C}}\right)$, contradicting the assumption that $\left(\mathbf{J}',\left(|J'_c|\right)_{c\in\mathbf{C}},\left(\text{Ind}(J'_c)\right)_{c\in\mathbf{C}}\right)$ is optimal. Therefore, if $(\mathbf{J}',\mathbf{z}'_1,\mathbf{z}'_0)$ is optimal for Eq. (11), this implies that $\mathbf{J}'$ is optimal for Eq. (10).

**Proof of (B):**

Suppose $\mathbf{J}$ optimizes Eq. (10). Consider any $(\mathbf{J}',\mathbf{z}'_1,\mathbf{z}'_0)$ that is optimal for Eq. (11). Using (A), we know that $\mathbf{J}'$ is then also optimal for Eq. (10). Therefore, Eq. (10) evaluated at $\mathbf{J}$ is equal to Eq. (10) evaluated at $\mathbf{J}'$. Using (L1) we know that $(\mathbf{J}',\mathbf{z}'_1,\mathbf{z}'_0) = \left(\mathbf{J}',\left(|J'_c|\right)_{c\in\mathbf{C}},\left(\text{Ind}(J'_c)\right)_{c\in\mathbf{C}}\right)$. Therefore Eq. (11) evaluated at $(\mathbf{J}',\mathbf{z}'_1,\mathbf{z}'_0)$ is equal to Eq. (10) evaluated at $\mathbf{J}'$. Note that Eq. (11) evaluated at $\left(\mathbf{J},\left(|J_c|\right)_{c\in\mathbf{C}},\left(\text{Ind}(J_c)\right)_{c\in\mathbf{C}}\right)$ is equal to Eq. (10) evaluated at $\mathbf{J}$. Therefore, $\left(\mathbf{J},\left(|J_c|\right)_{c\in\mathbf{C}},\left(\text{Ind}(J_c)\right)_{c\in\mathbf{C}}\right)$ is also an optimal solution to Eq. (11).

The equivalence between Eq. (10) and Eq. (11) is proved as a consequence of the proofs of (A) and (B). One consequence of this equivalence is that the constraints in Eq. (11) automatically ensure $z'_{1,c} = |J'_c|$ and $z'_{0,c} = \text{Ind}(J'_c)$, and therefore the solution spaces of Eq. (10) and Eq. (11) are equivalent.

**Comparison of the $L_0L_1$-norm formulation with the $L_1$-norm formulation**

In this section, we provide a mathematical explanation of why solutions to the $L_0L_1$-norm formulation may be closer to those of the ideal $L_0$-norm formulation than solutions to the $L_1$-norm formulation. Additionally, we also show that if the $L_1$-norm formulation provides an exact solution to the $L_0$-norm formulation (which is the case under certain conditions), then the $L_0L_1$-norm formulation also provides an exact solution to the $L_0$-norm formulation (although the solutions may be associated with different parameters).

From an intuitive perspective, the $L_0L_1$-norm formulation simply biases the solution more towards the $L_0$ norm, and so intuitively the solution associated with the $L_0L_1$-norm formulation is closer to that of $L_0$ norm for an equivalent $L_2$-norm error. This relationship can also be shown with formal arguments:

Let $J_1^*$ be an optimal solution to the $L_1$-norm formulation of Eq. (6) with regularization parameter $\mu_1$. There exists $\varepsilon_1$ such that $J_1^*$ is an optimal solution to

$$\begin{aligned} \min \ & \|\mathbf{J}\|_1 \\ s.t. \ & \left\|\mathbf{E}_{\text{DFT},S} - \mathbf{\Pi}_S \mathbf{J}\right\|_2^2 \leq \varepsilon_1 \end{aligned} \quad (12)$$

Now let $J_{10}^*$ be an optimal solution to the $L_0L_1$-norm formulation, Eq. (7), with regularization parameter $\mu_0'$ and $\mu_1'$. There exist $\varepsilon_{10}$ such that $J_{10}^*$ is an optimal solution to

$$\begin{aligned} \min \ & \|\mathbf{J}\|_1 + \frac{\mu_0'}{\mu_1'} \|\mathbf{J}\|_0 \\ s.t. \ & \left\|\mathbf{E}_{\text{DFT},S} - \mathbf{\Pi}_S \mathbf{J}\right\|_2^2 \leq \varepsilon_{10} \end{aligned} \quad (13)$$

In order to compare the proximity of the $L_0L_1$-norm formulation to the $L_0$-norm formulation, we need to have the same reference. Therefore, without loss of generality, we choose $\mu_1$, $\mu_0'$ and $\mu_1'$ such that

$$\varepsilon_1 = \varepsilon_{10} \equiv \varepsilon_0 \quad (14)$$

where $\varepsilon_0$ is defined as the common constraint parameter. We can now consider whether $J_1^*$ or $J_{10}^*$ is closer to the $L_0$-norm solution. To achieve this, we consider the $L_0$-norm problem

$$\begin{aligned} \min\ & \|\mathbf{J}\|_0 \\ s.t.\ & \|\mathbf{E}_{\text{DFT},S} - \mathbf{\Pi}_S \mathbf{J}\|_2^2 \leq \varepsilon_0 \end{aligned} \quad (15)$$

Note that $J_1^*$ and $J_{10}^*$ are both satisfying the constraint in Eq. (15) given (14).

We now show that $\|J_{10}^*\|_0 \leq \|J_1^*\|_0$ by contradiction:

Suppose $\|J_{10}^*\|_0 > \|J_1^*\|_0$. Since $J_1^*$ is an optimal solution to Eq. (12), we conclude $\|J_{10}^*\|_1 \geq \|J_1^*\|_1$. Therefore, $\|J_{10}^*\|_1 + \frac{\mu_0'}{\mu_1'}\|J_{10}^*\|_0 > \|J_1^*\|_1 + \frac{\mu_0'}{\mu_1'}\|J_1^*\|_0$, contradicting that $J_{10}^*$ is an optimal solution to Eq. (13). Hence, we have shown that

$$\|J_{10}^*\|_0 \leq \|J_1^*\|_0 \quad (16)$$

We conclude that given the same $L_2$-norm error, the $L_0L_1$-norm formulation provides a solution that is at least as good as the solution to the $L_1$-norm formulation assuming that the $L_0$-norm formulation is the targeted solution. Therefore, when R.I.P. is satisfied and the $L_1$-norm formulation provides the exact solution to the $L_0$-norm problem, the $L_0L_1$-norm formulation also provides an exact solution as the $L_0$-norm formulation given the same $L_2$-norm error.

### Hierarchical constraints

One advantage of the general $L_0L_1$ formulation of Eq. (11) is that it allows introducing additional constraints in a straightforward fashion. Here we demonstrate this concept by including an intuitive hierarchy constraint on the cluster basis. Similar concepts have previously been proposed in the literature [21,22].

One basic assumption of CE models is that *n*-body contributions to the configuration energy become less important the larger *n* becomes, so that the expansion can be truncated at a certain *n*. Often, 4-body interactions, *i.e.*, quadruplet clusters, are taken as the limit. To incorporate this principle in the model construction and to further reduce over-fitting, we introduce one more type of constraint into Eq. (11). This constraint is motivated by the natural hierarchy within the set of clusters $\mathbf{C}$. An example of this hierarchy is that a triplet is composed of three pairs (some of which may be equivalent due to symmetry), so the triplet has a higher order than any of these three pairs. If we consider some pair that is not contained within the triplet, we say that the pair and the triplet are incomparable. More generally, we define cluster *a* to have higher order than cluster *b* if cluster *a* contains cluster *b*. We further define the order of cluster *a* to be exactly one higher than the order of cluster *b*, if cluster *a* has higher order than cluster *b* and the size of cluster *a* is equal to the size of cluster *b* plus 1. The hierarchical set $H$ is defined as

$$H = \{(a,b) \in \mathbf{C} \times \mathbf{C} : \text{cluster } a \text{ is one order higher than cluster } b\}. \qquad (17)$$

Generally speaking, a higher order cluster can be thought of as a correction to lower order clusters when the lower order clusters are not sufficient to describe all interactions. To incorporate this physical intuition, we add the following type of constraints into Eq. (11):

$$z_{0,a} \leq z_{0,b} \quad \forall (a,b) \in H. \qquad (18)$$

Now, if $J_a \neq 0$, the original constraints in Eq. (11) lead to $z_{0,a} = 1$, and the additional constraint in Eq. (18) leads to $z_{0,b} = 1$.

We define any cluster *c* to be active if $z_{0,c} = 1$. Note that when the constraint of Eq. (18) is integrated into Eq. (11), a lower order cluster *b* automatically becomes active when a related higher order cluster *a* is active. Conversely, *b* being active no longer implies $J_b \neq 0$, but $J_b \neq 0$ does imply that *b* is active. In other words, when a higher order cluster *a* needs to be active, Eq. (18) requires that every sub-cluster *b* of cluster $a$ becomes also active, resulting in greater penalties to activate higher order clusters. As there are more higher order clusters than lower order clusters, Eq. (18) essentially guides the search of necessary higher order clusters in the $L_0L_1$-norm compressive sensing paradigm.

An illustration of the hierarchy constraint outlined above is shown in Figure 1 for a simple example lattice with four sites. The largest cluster in the example is a quadruplet cluster. When the ECI of that cluster becomes greater than zero, *i.e.*, when the quadruplet cluster becomes active, all smaller clusters also become active. This means, the $L_0$ regularization parameters of all lower-order clusters $c$ become $z_{0,c} = 1$, owing to the constraint Eq. (18).

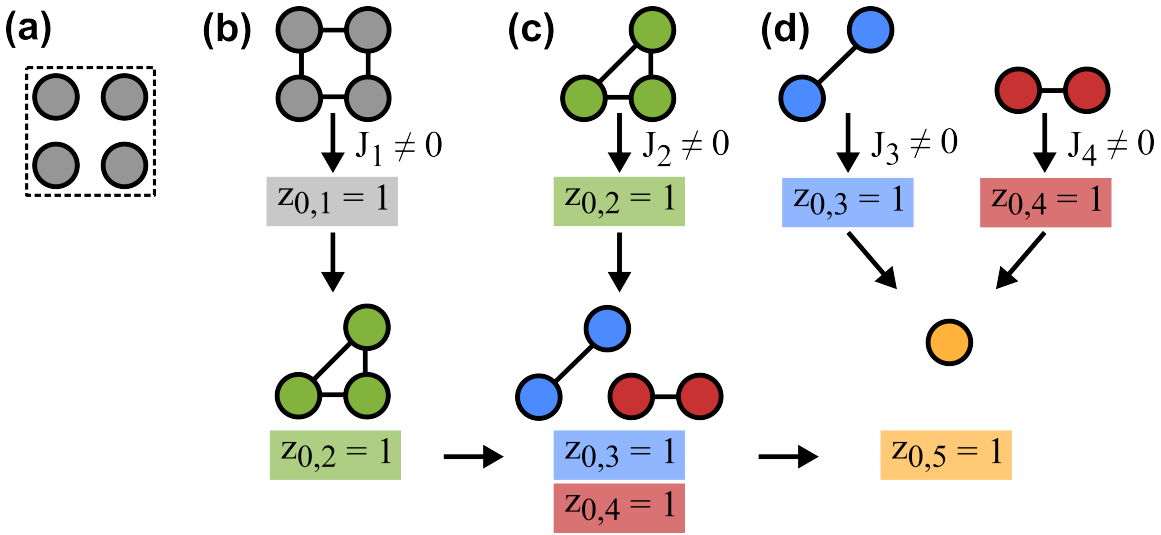

**Figure 1, cluster hierarchy constraint for an example lattice. (a) The example lattice with four equivalent sites. (b) If the ECI of the quadruplet cluster is different from zero, all smaller (lower-order) clusters become automatically active. (c) Activating the triplet cluster does not affect the quadruplet cluster as it is of higher order, but it results in activation of both pair clusters and the point cluster. (d) Activating any of the two pair clusters will also activate the point cluster.**

# Results

## Applications to cathode transition-metal oxide systems

Intercalation batteries function by reversible extraction and reintercalation of alkali metal or alkaline metal working ions from the cathode material, which typically is a transition-metal (TM) oxide or phosphate [46] The static ordering of the working ions (*e.g.*, $Li^+$, $Na^+$, $Mg^{2+}$) and the TM cations within the crystal structure of the cathode affect the achievable capacity and rate capability. Additionally, when working ions are extracted from the cathode, additional ordering interaction between the remaining working ions and vacancies can strongly affect the voltage profile and have to be considered in the computational design of novel battery materials. Owing to the large number of configurational degrees of freedom, determining the most stable working-ion/TM/vacancy orderings in realistic cathode materials at different states of charge is usually not feasible with DFT calculations and is a prototypical application of CE model simulations [47].

To benchmark our $L_0L_1$-norm compressive sensing CE approach, we apply the method to ordering problems in four different battery cathode systems: binary Li/Ti ordering in rocksalt-type $Li_xTi_{(1-x)}O$ [48,49], Mg/vacancy ordering in spinel-type $Mg_xCr_2O_4$ [50], Na/vacancy O3-type layered $Na_xCrO_2$ [51], and ternary Li/Ni/vacancy ordering in rocksalt- and spinel-type $Li_xNi_yO$ [52]. The different types of configurational ordering problems in the chosen materials provide a comprehensive benchmark of the new methodology.

All reference DFT energies of configurations used in the CE model construction were calculated under the Generalized Gradient Approximation with the Perdew-Burke-Ernzerhof (PBE) functional [53]. Further, a Hubbard-U correction [54,55] was employed for the *d* states of Ni and Cr using the U values from the work of Jain et al [56]. A plane wave basis set with an energy cutoff of 520 eV and projector-augmented wave (PAW) pseudopotentials were employed to describe the electronic wavefunctions [57,58]. All DFT calculations were done with the Vienna Ab initio

Simulation Package (VASP) [59,60]. For the chosen materials systems, initial sets of configurations were generated at various concentrations using an enumerating algorithm [61-63]. Subsequently, a ground-state searching algorithm for lattice models [16] was used to further expand the DFT dataset with additional low energy configurations. The geometry and cell parameters of all of the reference structures were optimized to nearby local minima. The final reference sets used for the CE model construction comprised 796 $Li_xTi_{(1-x)}O$ configurations, 249 $Mg_xCr_2O_4$ configurations, 281 $Na_xCrO_2$ configurations, and 453 $Li_xNi_yO$ configurations.

To assess the effectiveness of our $L_0L_1$-norm compressive sensing paradigm compared to the conventional $L_1$-norm compressive sensing for practical CE applications, we consider the 10-fold cross validation score (cv score) and the sparseness (model complexity) as a dual metric. The cv score is defined as the root mean square of the out-of-sample error. The average sparseness is the average $L_0$-norm of ECIs for each partition. In the case of the (conventional) $L_1$-norm compressive sensing, cross validation is performed with $\mu_1$ ranging from 0.001 to 10 in increments of 40 partitions on the logarithmic scale. For the $L_0L_1$-norm compressive sensing, cross validation is performed with $\mu_1$ ranging from 0.01 to 4 $eV/f.u.$, with 20 logarithmically-scaled increments and $\mu_0$ taking on 6 different values: $3 \times 10^{-5}$, $1 \times 10^{-4}$, $3 \times 10^{-4}$, $1 \times 10^{-3}$, $3 \times 10^{-3}$, and $1 \times 10^{-2}$ $eV/f.u.$. These ranges for $\mu_0$ and $\mu_1$ were empirically found to give the optimal results. Since the solution of the $L_0L_1$-norm formulation is not guaranteed to terminate in a polynomial time, the MIQP solver is terminated after 300 seconds for each partition, and the best result from within this time frame is extracted. For each $(\mu_0, \mu_1)$ pair, we record the average sparseness and the cv score.

To decide the current optimal CE fit, we define the *(sparseness, cv score)* front as follows: a *(sparseness, cv score)* pair is on the *(sparseness, cv score)* front if there exists no other *(sparseness', cv score')* such that *sparseness' ≤ sparseness* and *cv score' ≤ cv score*. In other words, any *(sparseness, cv score)* on the *(sparseness, cv score)* front is one optimal solution of the multi-objective minimization in the *(sparseness, cv score)* space. However, we emphasize that the standard choice of the final ECIs are those associated with the minimum cv score. Other ECIs are only chosen when there is a very strong preference for sparseness, a situation that has not occurred in our benchmark examples.

The *(sparseness, cv score)* front for the four different material systems is shown in Figure 2. We start our discussions with the comparison of the $L_0L_1$-norm formulation (illustrated with the blue curve) with conventional $L_1$-norm compressive sensing (illustrated with the red curve). For the $Li_xTi_{(1-x)}O$ system (Figure 2(a)), the optimal *(sparseness, cv score)* obtained with the conventional $L_1$-norm compressive sensing of Eq. (6) is *(129.7, 0.0346 eV/f.u.)*, while using the $L_0L_1$-norm formulation of Eq. (11) and Eq. (18) achieves an optimal *(sparseness, cv score)* of *(38.9, 0.0334 eV/f.u.)*. Hence, our new approach leads to a CE model that is 70% more sparse with similar predictive error. Further, based on the sparseness front shown in Figure 2(a), the $L_0L_1$-norm formulation still outperforms the standard $L_1$-norm compressive sensing when a higher cv score is traded in for a sparser CE. For a fixed *cv score*, the sparseness resulting from the $L_0L_1$-norm formulation is generally around 70% sparser. Conversely, for the same sparseness, the *cv score* is on average around 30% lower under the $L_0L_1$-norm compressive sensing paradigm. The above comparison indicates that the $L_0L_1$-norm formulation has much better compression capability, *i.e.*, its ability to extract important features from the DFT reference data.

For $Li_xNi_yO$, $Mg_xCr_2O_4$, and $Na_xCrO_2$, the results follow essentially the same trend with optimal CE models being 59%, 64%, and 66% sparser and 0.6%, 3.6%, and 0.7% more predictive in terms of *cv score*. The general trends and characteristics of the *(sparseness, cv score)* front are very similar for these materials, demonstrating the robustness and versatility of the $L_0L_1$-norm formulation.

We proceed to discuss the effect of hierarchical constraints, Eq. (18), on the $L_0L_1$-norm formulation. We compare the *(sparseness, cv score)* front for the $L_0L_1$-norm formulation with (blue curve in Figure 2) and without (black curve in Figure 2) hierarchical constraints. As seen in the figure, the two formulations show similar behavior. Both $L_0L_1$ formulations show substantial and comparable improvement of the sparseness over the $L_1$-norm compressive sensing of Eq. (6). For the $Li_xTi_{(1-x)}O$ system (Figure 2 (a)), the optimal *cv score* obtained with just the $L_0L_1$-norm formulation without hierarchical constraints is *0.0345 eV/f.u.*, while using the $L_0L_1$-norm formulation of Eq. (11) and Eq. (18) including the hierarchy constraints achieves an optimal *cv score* of *0.0334 eV/f.u.*. Therefore, the hierarchical constraints have resulted in a *cv score* improvement of 3.3% for this system. For the $Li_xNi_yO$ and $Na_xCrO_2$ systems, the optimal *cv scores* are similar. The inclusion of hierarchical constraints results in 0.4% and 0.5% *cv score* reduction respectively. However, for $Mg_xCr_2O_4$ the inclusion of hierarchical constraints leads to an increased *cv score* (-1.7% worse performance). We emphasize that the benefit of the $L_0L_1$-norm formulation is the capability to incorporate additional constraints, such as the physically intuitive hierarchy constraints, in a straightforward manner. This is however an entirely separate concern from the CE fit and it is well possible that intuitive constraints may lead to worse *cv scores* for specific systems though they might aid in the interpretation of the CE model. The remaining discussion in this paper assumes applications of the $L_0L_1$-norm formulation including hierarchical constraints.

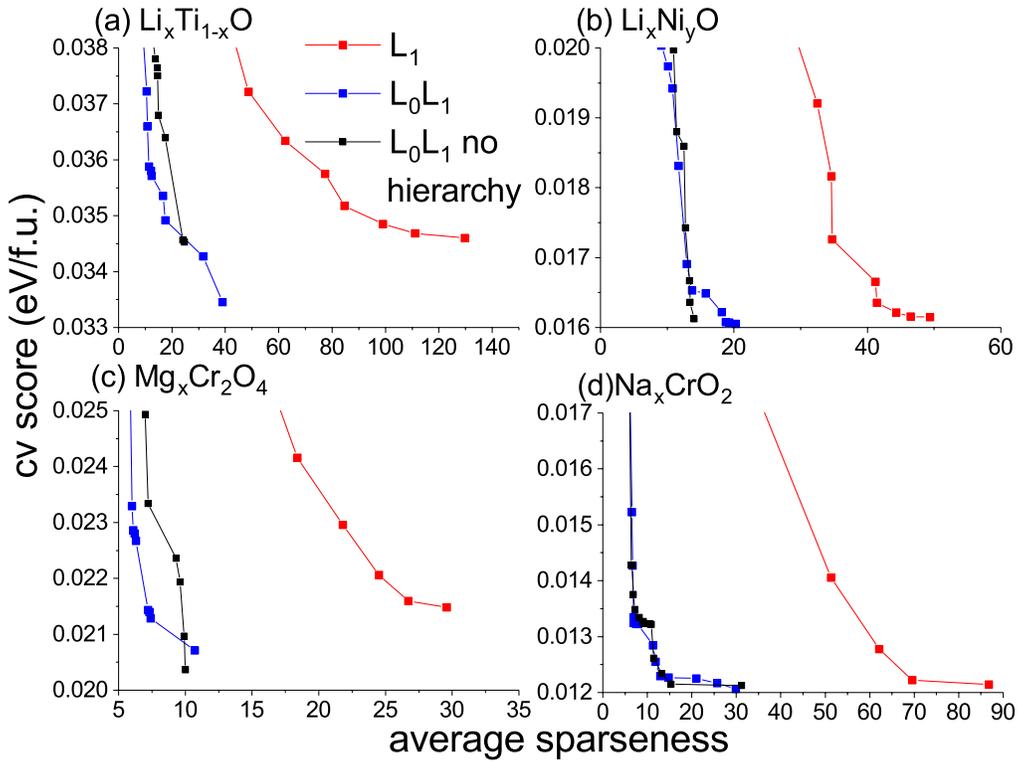

**Figure 2,** the *(sparseness, cv score)* **front as a measure of the quality of the CE fit for four different materials: (a) $Li_xTi_{(1-x)}O$, (b) $Li_xNi_yO$, (c) $Mg_xCr_2O_4$, and (d) $Na_xCrO_2$. The red line corresponds to the** *(sparseness, cv score)* **front constructed with standard $L_1$-norm compressive sensing, whereas the blue line corresponds to that constructed with the $L_0L_1$-norm paradigm** *and the black line corresponds to the L0L1-norm paradigm without the hierarchical constraints.*

To further analyze the sparseness improvements based on $L_0L_1$-norm formulation, we examine the differences in the ECI sets in detail for the example of the $Li_xTi_{(1-x)}O$ system. The ECIs resulting from both construction paradigms are plotted in Figure 3. The first eight ECIs (clusters 0 through 7) belong to the constant term, point term, and pair interactions. The triplet terms correspond to clusters 8 through 51, and the quadruplet terms are clusters 52 through 410. The vertical lines in Figure 3 indicate the limits of these three sets of ECIs. Clusters of the same order are sorted according to the maximal distance between two sites within them, which is an intuitive measure of the importance of the cluster.

Figure 3 (a) shows all ECIs for $Li_xTi_{(1-x)}O$. A first obvious difference between the two sets of ECIs is that there are only 7 nonzero ECIs beyond cluster 70 under the $L_0L_1$-norm formulation, while there are 96 under $L_1$-norm compressive sensing. Considering our initial assumption that the higher-order quadruplet terms are generally minor corrections to the pair and triplet interactions, one would expect that most of the corresponding ECIs should be 0. Clearly, from this physical perspective, the $L_0L_1$-norm formulation conforms better to the sparse physics paradigm. Figure 3 (b) shows only the ECIs up to cluster 70, which comprise all constant, point, pair, and triplet terms as well as the most important quadruplets with short internal distances. Interestingly, for the constant, point, and pair terms, the two sets of ECIs are very similar. The same is true for short distance triplets and quadruplets for which the $L_0L_1$-norm and $L_1$-norm formulations also generally agree with each other, indicating that both formulations capture the important short-range interactions well.

The key advantage of the $L_0L_1$-norm formulation is that higher-order clusters that correspond to longer ranged interactions are more compressed, in accordance with the expectations of the sparse physics paradigm. CE models with fewer clusters can be evaluated more efficiently, so that the sparser CE models result in a direct improvement of the computational efficiency. Additionally, the interpretation of the physics of a material is more straightforward for models with low complexity.

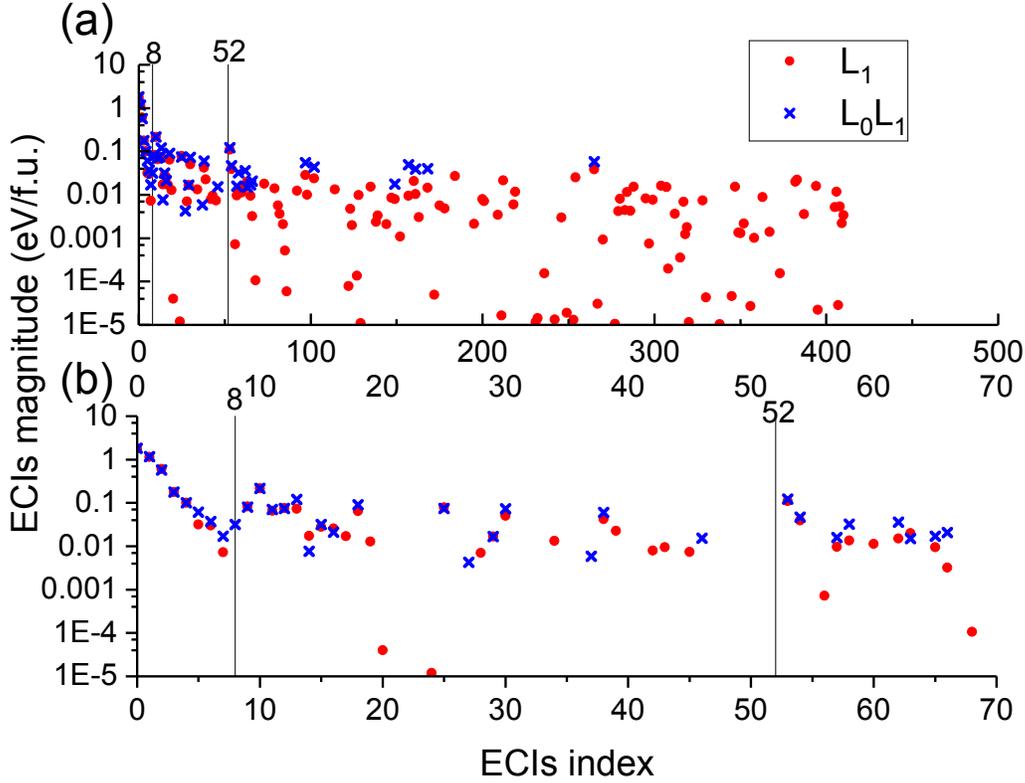

**Figure 3** Comparison of ECIs for $Li_xTi_{(1-x)}O$ obtained from conventional $L_1$-norm compressive sensing (red circles) and from our proposed $L_0L_1$-norm formulation (blue crosses). The vertical lines at index 8 and 52 indicate the transition from pairs to triplets and from triplets to quadruplets, respectively. (a) shows all of the ECIs, (b) shows only the ECIs up to index 70.

In summary, we have derived an $L_0L_1$-norm compressive sensing paradigm for the construction of cluster expansion lattice models from reference configurations that are not required to possess the restricted isometry property. We present an efficient implementation of this new paradigm using state-of-the-art mixed-integer quadratic programming techniques. We demonstrated the effectiveness and robustness of the method for four different transition-metal oxides with relevance as battery cathodes. The $L_0L_1$-norm compressive sensing formulation consistently results in significantly sparser CE models with predictive power as good or better than CE models constructed using conventional $L_1$-norm compressive sensing.

## Competing interests

The authors declare no competing financial interests.

## Acknowledgements

This work was primarily funded by the U.S. Department of Energy, Office of Science, Office of Basic Energy Sciences, Materials Sciences and Engineering Division under Contract No. DE-AC02-05-CH11231 (Materials Project program KC23MP). This work used the Extreme Science and Engineering Discovery Environment (XSEDE), which is supported by National Science Foundation grant no. ACI-1053575, and resources of the National Energy Research Scientific Computing Center (NERSC), a DOE Office of Science User Facility supported by the Office of Science of the US Department of Energy under contract no. DE-C02-05CH11231.